\documentclass[conference]{IEEEtran}
\IEEEoverridecommandlockouts
\newtheorem{theorem}{Theorem}
\usepackage{cite}
\usepackage{amsmath,amssymb,physics,tabularx,graphicx,multirow,hhline,algorithm,mathtools,algpseudocode,varwidth,diagbox,dsfont,graphicx}
\usepackage{textcomp}
\usepackage{xcolor}
\ifCLASSOPTIONcompsoc \usepackage[caption=false,font=normalsize,labelfont=sf,textfont=sf]{subfig}
\else
\usepackage[caption=false,font=footnotesize]{subfig}
\fi

\def\BibTeX{{\rm B\kern-.05em{\sc i\kern-.025em b}\kern-.08em
    T\kern-.1667em\lower.7ex\hbox{E}\kern-.125emX}}
\begin{document}

\title{A Holistic Framework for Parameter Coordination of Interconnected Microgrids against Disasters\\
}

\author{\IEEEauthorblockN{Tong Huang\IEEEauthorrefmark{1},
Hongbo Sun\IEEEauthorrefmark{2},
Kyeong Jin Kim\IEEEauthorrefmark{2},
Daniel Nikovski\IEEEauthorrefmark{2}, and
Le Xie\IEEEauthorrefmark{1}
\thanks{This work was mainly done while Tong Huang was working with Mitsubishi Electric Research Laboratories.}
}

\IEEEauthorblockA{\IEEEauthorrefmark{1}Department of Electrical and Computer Engineering\\
Texas A\&M University, College Station, Texas
\\\texttt{\{tonghuang,le.xie\}@tamu.edu}}
\IEEEauthorblockA{\IEEEauthorrefmark{2}
Mitsubishi Electric Research Laboratories, Cambridge, MA, 02139\\
\texttt{\{hongbosun,kkim,nikovski\}@merl.com}}
}

\maketitle
\begin{abstract}
This paper proposes a holistic framework for parameter coordination of a power electronic-interfaced microgrid interconnection against natural disasters. The paper identifies a transient stability issue in a microgrid interconnection. Based on recent advances in control theory, we design a framework that can systematically coordinate system parameters, such that post-disaster equilibrium points of microgrid interconnections are asymptotically stable. The core of the framework is a stability assessment algorithm using sum of squares programming. The efficacy of the proposed framework is tested in a four-microgrid interconnection. The proposed framework has potential to extend to microgrid interconnections with a wide range of hierarchical control schemes.
\end{abstract}


\section{Introduction}
The fragility of modern power grid under low-probability yet high-impact natural disasters has been exposed by extreme-weather-related blackouts during the past decade. Examples include 2011 Japan earthquake blackouts, 2012 Hurricane Sandy blackouts, and the electricity outage due to 2017 Hurricane Harvey, leading massive customers to loss their power supply \cite{7922501}. Since climate change will incur more extreme weather hazards and infrastructure aging exacerbates grid fragility, these massive blackouts are anticipated to occur more frequently in current electricity grid \cite{7105972}. Therefore, it is imperative to improve the resilience of power systems.

Microgrids have great potential for resilience enhancement thanks to their high operational flexibility \cite{7922501}. A microgrid has two operation modes, i.e., grid-connected mode and island mode. Under the normal operating condition, microgrids operate at grid-connected mode where control setpoints at points of common coupling (PCCs) are regulated by distribution system operators (DSO) to achieve global operation optima. When the main grid loses its desirable functions due to natural disasters, or when severe faults resulting from extreme weather hazards in some microgrids compromise normal operations of the main grid, some microgrids proactively enter the island mode \cite{7922501}. Based on these two operation modes of microgrids, a large body of literature discusses operation strategies for resilience enhancement. These strategies deal with critical issues in the time scale of quasi-steady state, such as microgrid sectionalization \cite{7017458}, restoration \cite{SunGM2019}, and resource allocation \cite{7105972}. However, steady-state studies do not consider microgrids' dynamical behavior at a finer time scale. As a result, the operation goal based on steady-state studies may not be achieved owing to insufficient examination of microgrids' stability. Therefore, it is equally critical to examine dynamic performance of networked microgrids in a much faster time scale in the presence of disasters.

There exists several efforts that scrutinize the dynamic performance of a single microgrid \cite{6155067} and interconnected microgrids \cite{7419922}. However, these efforts do not offer a guidance on how to systematically tune the system parameters to accomplish desirable dynamic performance. Also, the FACTS (Flexible Alternating Current Transmission System) techniques make it possible to modify topology parameters of interconnected microgrids. Although enhancing the flexibility of interconnected microgrids, such technical advances further complicate the parameters coordination problem by enlarging the search space. Systematically coordinating line and control parameters to achieve transient stability in the presence of natural disasters is insufficiently studied in the microgrid/resilience research communities. %Hence, it is necessary to develop a systematically parameter coordination framework for microgrids' dynamic performance improvement.

In view of the above challenges, we propose a holistic framework for parameter coordination of a power electronic-interfaced microgrid interconnection against natural disasters. The paper identifies a transient stability issue in a microgrid interconnection where one of microgrids enters islanding mode due to extreme weather hazards. Sum of squares programming \cite{HANCOCK2013960} is leveraged to design a stability assessment algorithm. Monte Carlo simulation embedded with the assessment framework is used for coordinating system parameters in a high-dimensional parameter space. The contributions of this paper are summarized are twofold. First, we propose a novel stability assessment algorithm. Compared with conventional stability assessment methods based on linear matrix inequalities \cite{7419922}, the proposed algorithm can characterize nonlinearity of microgrid interconnections more precisely, leading to its potential of offering a less conservative certificate of stability for microgrid interconnections. Second, the proposed framework can systematically tune system parameters, such that pre-designed equilibrium points of microgrid interconnections are asymptotically stable.
%The following two goals can be accomplished in the proposed framework: 1) the proposed framework can efficiently prioritize system parameters based on the parameters' significance in terms of system dynamics; 2) it can systematically tune the selected critical parameters, such that pre-designed equilibrium points of a microgrid interconnection can be achieved during natural disasters.

The rest of this paper is organized as follows: Section \ref{sec:math_des} describes the problems dealt in this paper rigorously; Section \ref{sec:framework} presents the proposed framework to coordinate parameters of interconnected microgrids; Section \ref{sec:Case_study} tests the framework using a four-microgrid interconnection; and Section \ref{sec:conclusion} concludes this paper.

\section{Mathematical Description of Interconnected Microgrids}
\label{sec:math_des}
\subsection{PCC Interface Dynamics and Network Constrains}
The PCC interface dynamics of microgrid $i$ can be characterized by the following differential equations \cite{7419922}:
\begin{subequations}\label{eq:MG_dynamics}
    \begin{align}
        T_{\text{a}i} \dot{\delta_i} + \delta_i - \delta_i^* &= D_{\text{a}i}(P_i^*-P_i)\\
        T_{\text{V}i} \dot{V_i} + V_i - V_i^* &= D_{\text{V}i}(Q_i^*-Q_i),
    \end{align}
\end{subequations}
where $V_i$, $\delta_i$, $P_i$ and $Q_i$ are the voltage magnitude, phase angle, real and reactive power injection at the $i$-th PCC, respectively; $V_i^*$, $\delta_i^*$, $P_i^*$ and $Q_i^*$ are the reference setting of $V_i$, $\delta_i$, $P_i$ and $Q_i$, respectively, which are dispatched by DSO according to steady state studies; $T_{\text{V}i}$ and $T_{\text{a}i}$ are the tracking time constants of voltage magnitude and phase angle, respectively; and $D_{\text{V}i}$ and $D_{\text{a}i}$ are droop gains of voltage magnitude and phase angle\cite{7419922}.

The PCC interface dynamics are coupled by the following power flow equations \cite{7419922}
\begin{subequations} \label{eq:power_flow}
    \begin{align}
        P_i &= V_i^2G_{ii}+ \sum_{k\ne i}V_iV_kY_{ik}\sin(\delta_i-\delta_k - \theta_{ik} + \pi/2),\\
        Q_i &= -V_i^2B_{ii}+ \sum_{k\ne i}V_iV_kY_{ik}\sin(\delta_i-\delta_k-\theta_{ik}), \forall i,
    \end{align}
\end{subequations}
where $G_{ii}$ and $B_{ii}$ are the self-conductance and self-susceptance of the $i$-th PCC; and $Y_{ik}\angle{\theta_{ik}}$ is the admittance of the branch from the $i$-th to $k$-th PCC, which can be used to express the resistance $R_{ik}$ and reactance $X_{ik}$ of the $i$-$k$ branch, viz.,
$R_{ik} = \cos(\theta_{ik})/Y_{ik}$, $ X_{ik} = -\sin(\theta_{ik})/Y_{ik}$.

We assume that there is a clear time-scale separation in the angle and voltage dynamics: the angle dynamics is much faster than the voltage magnitude dynamics such that voltage magnitudes can be approximated by their nominal values during the transient process \cite{7419922}. In this paper, we limit the research scope to angle dynamics of interconnected microgrids.

A $n$-microgrid interconnection can be described by a \emph{direct} graph $\mathcal{G}(\mathcal{V}, \mathcal{E})$, where $\mathcal{V} = \{1, 2, \ldots, n\}$ is the collection of $n$ buses; $\mathcal{E} = \{(i,k)\}$, whence ordered pair $(i,k)$ denotes the edge from bus $i$ to bus $k$. Note that $|\mathcal{E}|$ is the twice of the number of branches of the microgrid interconnection. Denote by $e_j:=(i,k)$ the $j$-th element in $\mathcal{E}$. The dynamics of the $i$-th microgrid can be expressed in the following form:
\begin{subequations}
    \begin{align}
        &T_{\text{a}i} \Delta \dot{\delta_i} + \Delta\delta_i = -D_{\text{a}i}\sum_{e_j \in \mathcal{E}_i}\kappa_{e_j}\phi_{e_j}(y_{e_j}),\\
        &\phi(y_{e_j}) = \sin(y_{e_j}+y_{e_j}^*)-\sin(y_{e_j}^*) \quad \forall e_j \in \mathcal{E}_i,\label{eq:Lure_system_nonlinear}
    \end{align}
\end{subequations}
where $\Delta \delta_i = \delta_i - \delta_i^*$; $e_j = (i,k)\in \mathcal{E}_i$, $y_{e_j} = \Delta \delta_i - \Delta \delta_k$, $y_{e_j }^*=\delta_i^*-\delta_k^*+\pi/2 -\theta_{ik}$, and $\kappa_{e_j}=V_i^*V_k^*Y_{ik}$ for all $e_j$; and 
\begin{equation*}
    \begin{aligned}
        \mathcal{E}_i:=\{(i,k)|&\text{$k$ is the index of the first}
        \\&\text{neighbor of microgrid $i$}.\}\subseteq \mathcal{E}.
    \end{aligned}
\end{equation*}

Define matrices $X=[x_{p,q}]\in \mathbb{R}^{n\times |\mathcal{E}|}$ and $K = \text{diag}(\kappa_{e_1},\kappa_{e_2},\ldots, \kappa_{e_{|\mathcal{E}|}})$, where
\begin{equation}
x_{p,q} =
    \begin{cases}
    1 & \forall p = 1, 2, \ldots,n \wedge (p,q)\in\mathcal{E}_p,\\
    0 & \text{otherwise}.
    \end{cases}
\end{equation}
The dynamics of the $n$-interconnected microgrids can be characterized by the following state-space form \cite{8116598,8345676,huang2018synchrophasor,8362302}
\begin{subequations}\label{eq: state_space}
    \begin{align}
        &\dot{\boldsymbol{\delta}}_n = A_n\boldsymbol{\delta}_n+B_n\boldsymbol{\phi}(\mathbf{y}_n),\\
        &\mathbf{y}_n = C_n \boldsymbol{\delta}_n,
    \end{align}
\end{subequations}
where $C_n\in \mathbb{R}^{ |\mathcal{E}|\times n}$ is the connectivity matrix;
\begin{subequations} \label{eq:compact_form}
    \begin{align}
        &\boldsymbol{\delta}_n=[\Delta\delta_1, \Delta\delta_2, \ldots, \Delta\delta_n]^{\top};\\
        &\mathbf{y}_n = [y_{e_1},y_{e_2},\ldots,y_{e_{|\mathcal{E}|}}]^{\top};\\
        &A_n = \text{diag}(-1/T_{\text{a}1}, -1/T_{\text{a}2}, \ldots, -1/T_{\text{a}n});\\
        &\begin{aligned}
        &B_n = DXK,\\ 
        &D=\text{diag}(-D_{\text{a}1}/T_{\text{a}1}, -D_{\text{a}2}/T_{\text{a}2}, \ldots, -D_{\text{a}n}/T_{\text{a}n});    
        \end{aligned}\\
        \text{and } &\boldsymbol{\phi}(\mathbf{y}_n) = [\phi_{e_1}(y_{e_1}), \phi_{e_2}(y_{e_2}),\ldots,\phi_{e_{|\mathcal{E}|}}(y_{e_{|\mathcal{E}|}})]^{\top}.
    \end{align}
\end{subequations}
Note that $\boldsymbol{\phi}(\mathbf{y}_n)$ introduces nonlinearities and the origin is the equilibrium point of the dynamical system \eqref{eq:compact_form}.

Set $\mathcal{E}$ can be partitioned into two subsets $\mathcal{E}^0=\{\mathcal{E}_i^0\}$ and $\mathcal{E}^1$, where $\mathcal{E}_i^0 := \{(i,k)|(i,k)\in\mathcal{E}\wedge i\le k\}$, and $\mathcal{E}^1$ can be obtained by swapping bus numbers in each element in $\mathcal{E}^0$. $\mathbf{y}'_n:=[y_{e_1^0},y_{e_2^0},\ldots,y_{e_{|\mathcal{E}^0|}^0}]^{\top}$. Suppose that resistances of interconnection lines are zero, then
\begin{equation}
    \phi(y_{e_j^0}) = -\phi(y_{e_j^1}).
\end{equation}
The input vector $\boldsymbol{\phi}(\mathbf{y})$ can be written as
\begin{equation}
    \boldsymbol{\phi}(\mathbf{y}) = 
    \begin{bmatrix}
        \boldsymbol{\phi}_0(\mathbf{y}')\\
        -\boldsymbol{\phi}_0(\mathbf{y}')
    \end{bmatrix}
    = F\boldsymbol{\phi}_0(\mathbf{y}')
\end{equation}
where $\boldsymbol{\phi}_0 = [\phi(y_{e_1^0}),\phi(y_{e_2^0}),\ldots, \phi(y_{e_{|\mathcal{E}^0|}^0})]^{\top}$, and $F=[I_{|\mathcal{E}^0|}, -I_{|\mathcal{E}^0|}]^{\top}\in \mathbb{R}^{|\mathcal{E}|\times |\mathcal{E}_0|}$, whence $I_{|\mathcal{E}|}$ denotes a $|\mathcal{E}|$ by $|\mathcal{E}|$ identity matrix.
By replacing $\boldsymbol{\phi}$ with $\boldsymbol{\phi}_0$ in \eqref{eq: state_space}, a microgrid interconnection with lossless tie lines can be characterized by
\begin{subequations}\label{eq: state_space_prime}
    \begin{align}
        &\dot{\boldsymbol{\delta}}_n = A_n\boldsymbol{\delta}_n+B'_n\boldsymbol{\phi}_0(\mathbf{y}_n'),\\
        &\mathbf{y}_n' = C'_n \boldsymbol{\delta}_n,
    \end{align}
\end{subequations}
where $B'_n = DXKF$. Note that the number of nonlinearities in \eqref{eq: state_space_prime} is half of that in \eqref{eq: state_space}.

\subsection{Stability Issues Due to Natural Disasters}
Without loss of generality, we assume that a natural disaster has severe impact on the $(n+1)$-th to $m$-th microgrids, where integer $m>n$, such that the $(m-n)$ microgrids enter the islanding mode. The locations of the impacted microgrids are assumed to be reliably predicted by weather forecast. Before the $(m-n)$ microgrids enter the islanding mode, the dynamic behavior of the $n$ remaining interconnected microgrids can be described by \eqref{eq: state_space_prime} with a pre-designed equilibrium point $\boldsymbol{\delta}_n^*$, where $\delta_i=\delta_i^*$ for all $i=1,2,\ldots, n$.

\begin{figure}
    \centering
    \includegraphics[width = 3in]{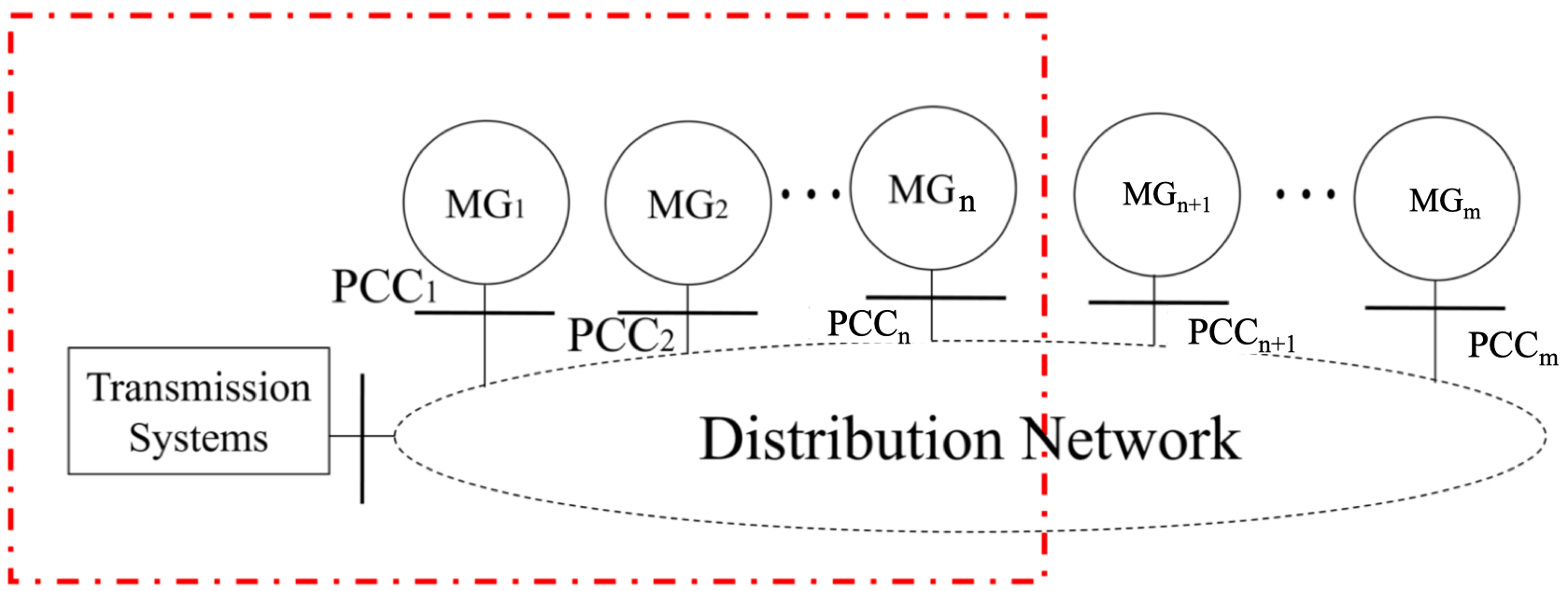}
    \caption{$m$ interconnected microgrids}
    \label{fig:N_Microgrids}
\end{figure}

Let vector $\boldsymbol{\alpha}$ collects all tunable parameters, such as control parameters $T_{\text{a}i}$, $D_{\text{a}i}$ and line parameters.
DSO expects that, after the $(m-n)$ microgrids are isolated, the remaining $m$ microgrids (in the red box in Fig. \ref{fig:N_Microgrids}) can reach a pre-designed equilibrium point $\boldsymbol{\delta}^*_n$. However, the nonlinear system described by \eqref{eq: state_space_prime} with parameter vector $\boldsymbol{\alpha}$ may not be configured in a manner that the pre-designed equilibrium $\boldsymbol{\delta}^*_n$ is asymptotically stable in the large. Therefore, the $n$ interconnected microgrids might not be stable or they may end up with an undesirable operation condition where physical constrains of the grid are severely violated. Under such a condition, a key question is \emph{how to tune parameters $\boldsymbol{\alpha}$, such that the pre-designed equilibrium} $\boldsymbol{\delta}^*_n$ \emph{ is asymptotically stable in the large after the $(m-n)$ microgrids impacted by natural disasters enter the islanding mode}. In Section \ref{sec:Case_study}-A, a numerical example is employed to demonstrate the stability issue described above.
\section{Framework for Parameter Coordination}
\label{sec:framework}
\subsection{Stability Assessment Based on Sum of Squares Programming} % (fold)
\label{sub:stability_assessment_based_on_sum_of_square}
\subsubsection{Generalised Sector for Nonlinearities} % (fold)
\label{ssub:generalised_sector_for_nonlinearities}
The nonlinearities in interconnected microgrids \eqref{eq: state_space_prime} are introduced by $\boldsymbol{\phi}_0(\mathbf{y}_n')$. Each element in $\boldsymbol{\phi}_0(\mathbf{y}_n')$ can be expressed by \eqref{eq:Lure_system_nonlinear}. In what follows, we aim to use a polynomial inequality to bound \eqref{eq:Lure_system_nonlinear}. 

\begin{figure}[thb]
    \centering
    \includegraphics[width = 2.5in]{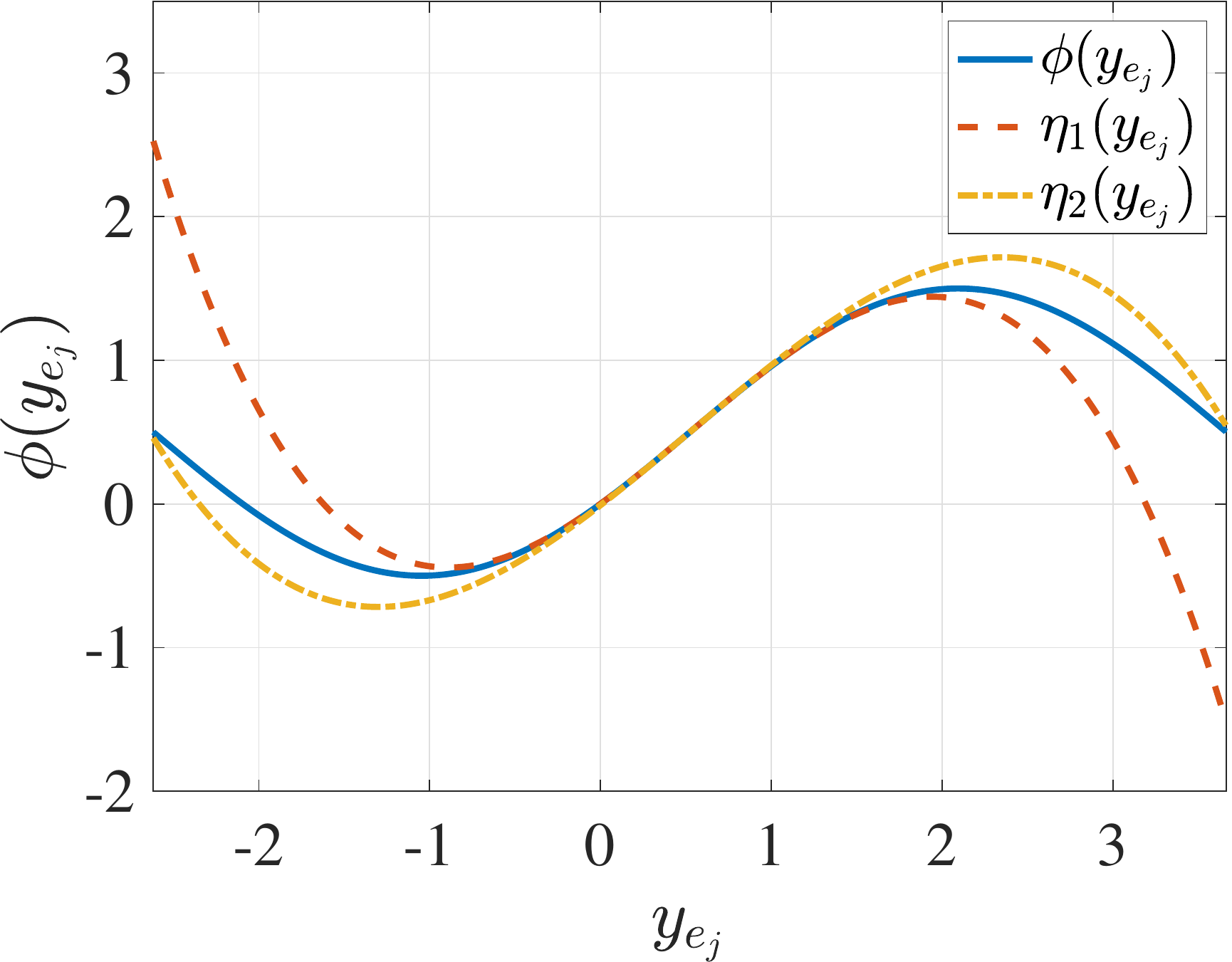}
    \caption{Nonlinearity $\phi(y_{e_j})$ with $y_{e_j}^*=-\pi/6$ in $[-5\pi/6, 7\pi/6]$.}
    \label{fig:eq:sector_example}
\end{figure}

For $-\pi\le y\le \pi$, $\nu(y)=\sin(y)$ is in a generalised sector \cite{HANCOCK2013960}
\begin{equation*}
    \left(\nu-y+\frac{y^3}{6}\right)\left(\nu-y+\frac{y^3}{10}\right)\le 0.
\end{equation*}
Then, for $y_{e_j}\in[-\pi-y_{e_j}^*, \pi-y_{e_j}^*]$, $\phi(y_{e_j})$ in \eqref{eq:Lure_system_nonlinear} satisfy
\begin{equation}\label{eq:generalised_sector}
    \left[\phi_{e_j} - \eta_1(y_{e_j})\right]\left[\phi_{e_j}-\eta_2(y_{e_j})\right]\le 0,
\end{equation}
where
\begin{subequations}
    \begin{align}
        \eta_1(y_{e_j}) &= (y_{e_j}+y_{e_j}^*)-\frac{1}{6}(y_{e_j}+y_{e_j}^*)^3-\sin(y_{e_j}^*),\\
        \eta_2(y_{e_j}) &= (y_{e_j}+y_{e_j}^*)-\frac{1}{10}(y_{e_j}+y_{e_j}^*)^3-\sin(y_{e_j}^*).
    \end{align}
\end{subequations}
Figure \ref{fig:eq:sector_example} shows an example where the nonlinearity $\phi(y_{e_j})$ with $y_{e_j}^*=-\pi/6$ is bounded by polynomials $\eta_1(y_{e_j})$ and $\eta_2(y_{e_j})$ in $[-5\pi/6, 7\pi/6]$. It is worth noting that, for the purpose of bounding sinusoidal-type of nonlinearities in certain regions, the generalised sector \eqref{eq:generalised_sector} based on polynomials are much tighter than the linear sectors proposed in \cite{7419922}.

The left-hand-side of \eqref{eq:generalised_sector} is a function of $\phi_{e_j}$ and $y_{e_j}$. Denote such a function by $r_j(\phi_{e_j},y_{e_j})$. The feasible range of inequality \eqref{eq:generalised_sector} is the solution to the following inequality in terms of $y_{e_j}$:
\begin{equation}\label{eq:feasible_range}
    (y_{e_j}+\pi+y_{e_j}^*)(y_{e_j}-\pi+y_{e_j}^*)\le 0.
\end{equation}
Owing to $y_{e_j}$ is a function of $\boldsymbol{\delta}_n$, the left-hand-side of \eqref{eq:feasible_range} is a function of $\boldsymbol{\delta}_n$ which is represented by $a_j(\boldsymbol{\delta}_n)$. With a domain 
\begin{equation} \label{eq:generalised_sector_a}
    \{\boldsymbol{\delta}_n\in\mathbb{R}^n|\mathbf{a}(\boldsymbol{\delta}_n) \preccurlyeq 0\},
\end{equation}
the generalised sector for $\boldsymbol{\phi}_0(\mathbf{y'}_n)$ can be expressed compactly by
\begin{equation} \label{eq:generalised_sector_r}
    \mathbf{r}(\boldsymbol{\phi}_0, \boldsymbol{\delta}_n) \preccurlyeq 0,
\end{equation}
where 
\begin{equation} \label{eq:a_vector}
    \mathbf{a}(\boldsymbol{\delta}_n):=[a_1(\boldsymbol{\delta}_n), \ldots,a_{|\mathcal{E}^0|}(\boldsymbol{\delta}_n)]^{\top};
\end{equation}
\begin{equation} \label{eq:r_vector}
    \mathbf{r}(\boldsymbol{\phi}_0, \boldsymbol{\delta}_n):=[r_1(\phi_{e_1},\boldsymbol{\delta}_n),\ldots,r_{|\mathcal{E}^0|}(\phi_{e_{|\mathcal{E}^0|}})]^{\top};
\end{equation}
and ``$\preccurlyeq$'' denotes element-wised ``less than or equal to''. In sum, the dynamics of a microgrid interconnection are described by \eqref{eq: state_space_prime} with nonlinearity bounded by the generalised sector \eqref{eq:generalised_sector_r} in the domain \eqref{eq:generalised_sector_a}.

% subsubsection generalised_sector_for_nonlinearities (end)

\subsubsection{A SOS-based Stability Assessment Algorithm} % (fold)
\label{ssub:stability_assessment_algorithm}
A theorem developed in \cite{HANCOCK2013960} is leveraged to assess the asymptotic stability of the microgrid interconnection described by \eqref{eq: state_space_prime}. Before introducing the theorem for asymptotic stability assessment, we present a notation relevant to polynomials and the definition of sum of squares (SOS).

For vectors $\mathbf{x}_1$ and $\mathbf{x}_2$, $\boldsymbol{\psi}(\mathbf{x}_1,\mathbf{x}_2)\in \mathbb{R}^q[\mathbf{x}_1,\mathbf{x}_2]$ denotes $\boldsymbol{\psi}(\mathbf{x}_1,\mathbf{x}_2)$ is a $q$-dimensional vector of polynomials in $\mathbf{x}_1$ and $\mathbf{x}_2$. For example, $\mathbf{a}(\boldsymbol{\delta}_n)\in \mathbb{R}^{|\mathcal{E}^0|}[\boldsymbol{\delta}_n]$, and $\mathbf{r}(\boldsymbol{\phi}_0,\boldsymbol{\delta}_n)\in \mathbb{R}^{|\mathcal{E}^0|}[\boldsymbol{\phi}_0,\boldsymbol{\delta}_n]$. $\boldsymbol{\psi}(\mathbf{x}_1,\mathbf{x}_2)$ is \emph{SOS} if each polynomial $\psi_i(\mathbf{x}_1,\mathbf{x}_2)$ in polynomial vector $\boldsymbol{\psi}$ can be expressed as SOS polynomials in $\mathbf{x}_1$ and $\mathbf{x}_2$, i.e.,
\begin{equation}
    \psi_i(\mathbf{x}_1,\mathbf{x}_2) = \sum_{k=1}^{w_i} h_{ik}(\mathbf{x}_1,\mathbf{x}_2)^2 \quad \forall i = 1,2, \ldots,q,
\end{equation}
where $h_{ik}(\mathbf{x}_1,\mathbf{x}_2)$ is a polynomial in $(\mathbf{x}_1,\mathbf{x}_2)$.
\begin{theorem} \label{thm:SOSThm}
    The equilibrium point of the system \eqref{eq: state_space_prime} with $\boldsymbol{\phi}_0$ bounded by \eqref{eq:generalised_sector_r} in the domain \eqref{eq:generalised_sector_a} is asymptotically stable in a finite domain, if there exists a polynomial $V(\boldsymbol{\delta}_n)$ with $V(\mathbf{0})=0$, SOS polynomials $\mathbf{s}_1, \mathbf{s}_2\in \mathbb{R}^{|\mathcal{E}^0|}[\boldsymbol{\phi}_0,\boldsymbol{\delta}_n]$, and strictly positive definite polynomials $\sigma_1(\boldsymbol{\delta}_n),\sigma_2(\boldsymbol{\delta}_n)$, such that 
\begin{equation}
    V(\boldsymbol{\delta}_n)-\sigma_1(\boldsymbol{\delta}_n) \quad \text{is SOS in $\boldsymbol{\delta}$},
    \label{eq:SOScondition1}
\end{equation}
\begin{equation}\label{eq:SOScondition2}
	\begin{aligned}
		-\nabla V(\boldsymbol{\delta}_n)&[A_n\boldsymbol{\delta}_n+B_n\boldsymbol{\phi}_0]-\sigma_2(\boldsymbol{\delta}_n)
    + \mathbf{s}_1(\boldsymbol{\phi}_0,\boldsymbol{\delta}_n)^{\top}\mathbf{r}(\boldsymbol{\phi}_0,\boldsymbol{\delta}_n)\\
    &+\mathbf{s}_2(\boldsymbol{\phi}_0,\boldsymbol{\delta}_n)^{\top}\mathbf{a}(\boldsymbol{\delta}_n)\quad \text{is SOS in $(\boldsymbol{\delta}_n,\boldsymbol{\phi}_0)$}.
	\end{aligned}
\end{equation}
\end{theorem}

In Theorem \ref{thm:SOSThm}, the degree of polynomials/polynomial vectors $V$, $\mathbf{s}_1$, $\mathbf{s}_2$, $\sigma_1$, and $\sigma_2$ are user-defined parameters, which are represented by $l_V$, $l_{\mathbf{s}1}$, $l_{\mathbf{s}2}$, $l_{\sigma_1}$, and $l_{\sigma_2}$, respectively. Whether polynomials are SOS can be checked by SOSTOOLS \cite{sostools}. Based on Theorem \ref{thm:SOSThm}, Algorithm \ref{alg:1} is proposed for the purpose of assessing asymptotic stability of a pre-designed equilibrium point of a given microgrid interconnection.

\begin{algorithm}
    \caption{Stability Assessment Algorithm Based on SOS} \label{alg:1}
        \begin{algorithmic}[1]
            \Function{\tt StablityAssess}{$l_V$, $l_{\mathbf{s}1}$, $l_{\mathbf{s}2}$, $l_{\sigma_1}$, $l_{\sigma_2}$, $\boldsymbol{\alpha}$}
            \State Construct $A_n'$, $B_n'$, and $C_n'$ based on $\boldsymbol{\alpha}$, \eqref{eq: state_space}, \eqref{eq:compact_form}, \eqref{eq: state_space_prime};
            \State Construct $\mathbf{r}$ based on \eqref{eq:generalised_sector} and \eqref{eq:r_vector};
            \State Construct $\mathbf{a}$ based on \eqref{eq:feasible_range} and \eqref{eq:a_vector};
            \State Check the feasibiltiy of \eqref{eq:SOScondition1}, \eqref{eq:SOScondition2} in SOSTOOLS;
            \If{\eqref{eq:SOScondition1} and \eqref{eq:SOScondition2} are feasible}
            \State $\zeta = 1$;
            \Else
            \State $\zeta = 0$;
            \EndIf
            \State\textbf{return}: $\zeta$.
            \EndFunction
        \end{algorithmic}
\end{algorithm}
% subsubsection stability_assessment_algorithm (end)

\subsection{Systematic Parameter Modification}
This subsection proposes an algorithm for systematic parameter modification. Denote by $\alpha_i$ the $i$-th entry in vector $\boldsymbol{\alpha}$. Set $\mathcal{I}$ collects the indices of all adjustable parameters. Each adjustable parameter $\alpha_i$ for $i\in \mathcal{I}$ has a upper and lower bound, represented by $\gamma_i'$ and $\gamma_i''$, respectively. Denote by $\boldsymbol{\alpha}'= [\alpha'_i]$ the randomized version of $\boldsymbol{\alpha}$, where
\begin{equation}
    \alpha_i' = \begin{cases}
    \gamma_i \alpha_i & i\in \mathcal{I}\\
    \alpha_i & \text{otherwise},
    \end{cases}
    \label{eq:randomized}
\end{equation}
in which $\gamma_i$ is a realization of random variable $\Gamma_i$ which has a uniform distribution, i.e., $\Gamma_i \sim \mathcal{U}(\gamma_i', \gamma_i'')$. With the above notations, the procedure for parameter modification is described in Algorithm \ref{alg:2}, where Monte-Carlo simulation time $N$ is defined by users; $\|\cdot\|_2$ is the $\mathcal{L}$-$2$ norm; $\boldsymbol{\Gamma}'= \{\gamma_i'|i\in\mathcal{I}\}$; and $\boldsymbol{\Gamma}''= \{\gamma_i''|i\in\mathcal{I}\}$. 

Given adjustable parameters ($\boldsymbol{\alpha}, \mathcal{I}$) associated with their tunable ranges ($\Gamma', \Gamma''$), Algorithm \ref{alg:2} first searches for parameter combinations $\mathcal{S}$ that enable all states to stay close to the pre-designed equilibrium $\mathbf{o}_m'$ (Line $3$ to $9$ in Algorithm \ref{alg:2}). It is worth noting that, as $\mathcal{I}$ may include topology parameter indices, the post-disaster equilibrium point $\mathbf{o}_m'$ should be revised accordingly. Then, Algorithm \ref{alg:2} returns one parameters combination $\mathbf{v}^*$ that minimizes the Euclidean distance from the initial parameters $\boldsymbol{\alpha}$ among all eligible combinations $\mathcal{S}$, as well as the corresponding equilibrium point $\mathbf{o}_m^*$.

\begin{algorithm}
    \caption{Systematic Parameter Modification} \label{alg:2}
        \begin{algorithmic}[1]
            \Function{\tt ParaMod}{$N,\boldsymbol{\alpha},\mathcal{I},l_V, l_{\mathbf{s}1}, l_{\mathbf{s}2}, l_{\sigma_1}, l_{\sigma_2}, \boldsymbol{\Gamma}', \boldsymbol{\Gamma}''$}
                \State $\mathcal{S} \gets \O; \mathbf{o}_m^* \gets 0;$
                \While{$k = 1, 2, \ldots, N$}
                    \State Construct $\boldsymbol{\alpha}'$ via \eqref{eq:randomized};
                    \State Update $\mathbf{o}_m'$ via power flow studies based on $\boldsymbol{\alpha}';$
                    \State $\xi\gets$\texttt{ StablityAssess}($l_V$, $l_{\mathbf{s}1}$, $l_{\mathbf{s}2}$, $l_{\sigma_1}$, $l_{\sigma_2}$, $\boldsymbol{\alpha}$);
                    \If{$\xi = 1$}
                        \State $\mathcal{S}\gets \mathcal{S}\cup \boldsymbol{\alpha}'$;
                    \EndIf
                \EndWhile
                \If{$\mathcal{S}=\O$}
                \State $\mathbf{v}^*\gets0$;
                \Else
                \State $\mathbf{v}^* \gets \text{arg}\min_{\mathbf{v}\in \mathcal{S}}\|\boldsymbol{\alpha}-\mathbf{v}\|_2$;
                \State Update $\mathbf{o}_m'$ via power flow studies based on $\mathbf{v}^*$;
                \State $\mathbf{o}_m^* \gets \mathbf{o}_m'$
                \EndIf
                \State \Return $\mathbf{v}^*, \mathbf{o}_m^*$.
            \EndFunction
        \end{algorithmic}
\end{algorithm}

% subsection stability_assessment_based_on_sum_of_square (end)
\section{Case Study} \label{sec:Case_study}
This section leverages a four-microgrid interconnection (Fig. \ref{fig: Four_MG_Motivation}-a) to validate the proposed framework. We first identify the stability issue in the test system after one microgrid enters the islanding mode. Then, the proposed framework based on SOS is employed for coordinating system parameters.
\subsection{Motivating Example} % (fold)
\label{sub:motivating_example}

\begin{figure}[h]
\centering
\subfloat[]{\includegraphics[width=1.6in]{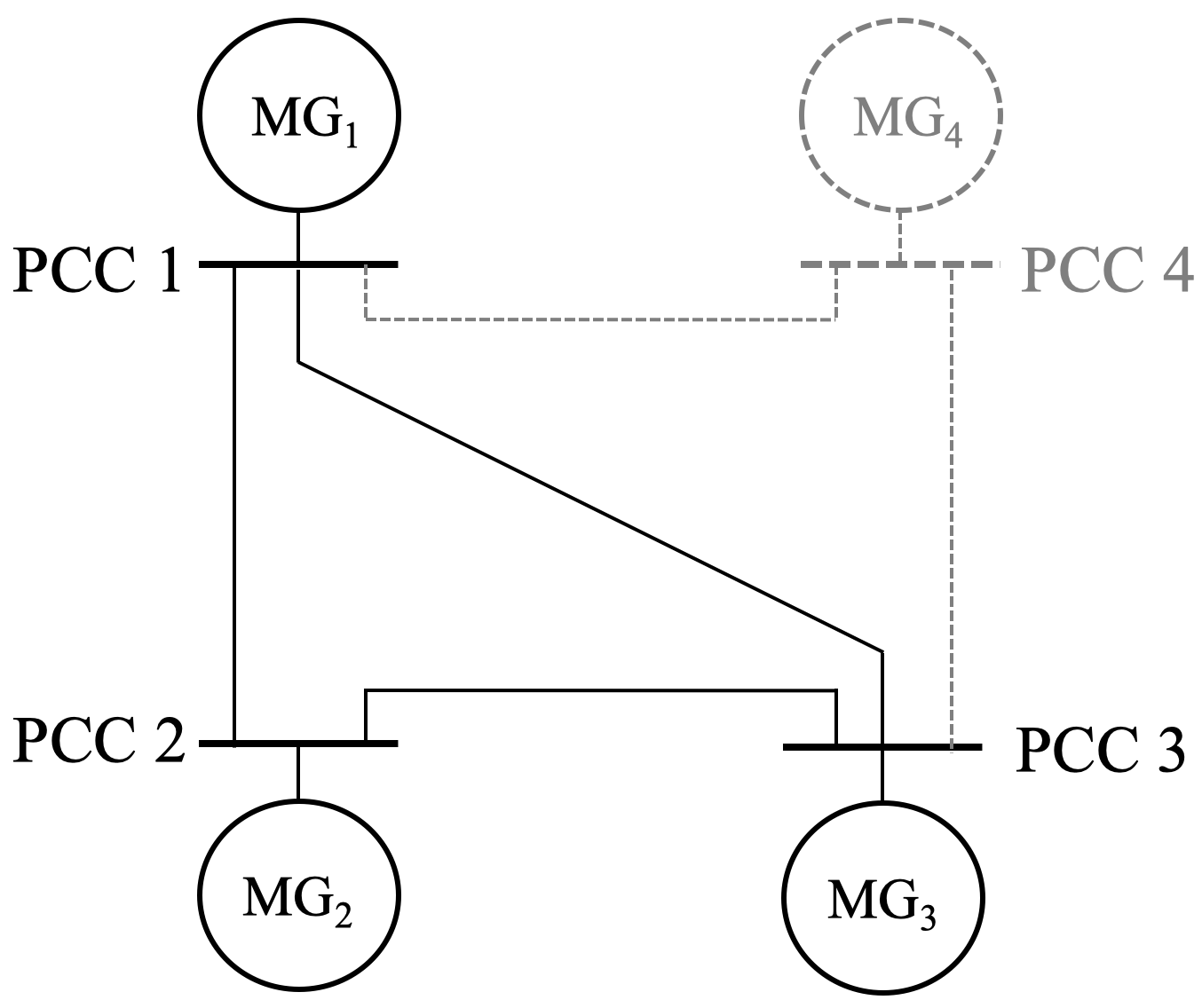}} 
\hfil
\subfloat[]{\includegraphics[width=1.7in]{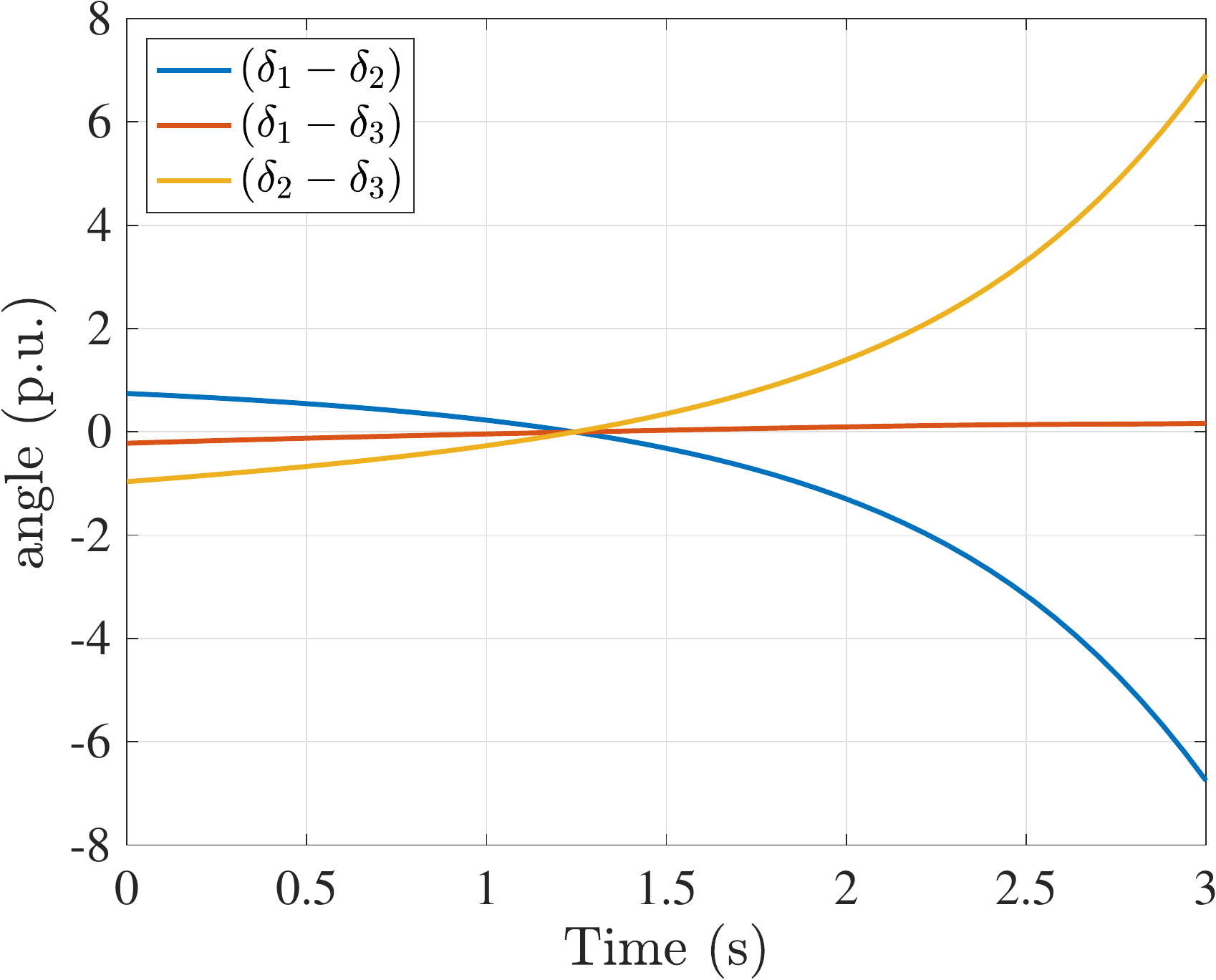}}
\hfil
\caption{(a) A four-microgrid interconnection where microgrid $4$ enters the islanding mode; (b) The evolution of voltage angle differences after the microgrid $4$ enters the islanding mode.}
\label{fig: Four_MG_Motivation}
\end{figure}
Suppose that a disaster is predicted to happened at Microgrid (MG) $4$, forcing MG $4$ to enter the islanding mode. The reactances of lines $1$-$2$, $1$-$3$ and $2$-$3$ are $0.45$, $0.65$, and $0.66$ in per unit (p.u.), respectively. The control parameter vectors $[T_{\text{a}1}, T_{\text{a}2},T_{\text{a}3}]^{\top}=[4.10, -0.78, 2.56]^{\top}$ and $[D_{\text{a}1}, D_{\text{a}2},D_{\text{a}3}]^{\top}=[0.0286, -0.0178, -0.0284]^{\top}$. According to economic/safety-based steady-state study, the post-disaster steady states are assigned at
\begin{subequations}
    \begin{align*}
        &[V_1^*, V_2^*, V_3^*]^{\top} = [1, 1.05, 0.95]^{\top} \text{(in p.u.)},\\
        &[\delta_1^*, \delta_2^*, \delta_3^*]^{\top} = [0, -0.57, -0.24]^{\top} \text{(in rad.)}.
    \end{align*}
\end{subequations}

However, the transient study (Fig. \ref{fig: Four_MG_Motivation}-b) suggests that the pre-designed steady states (equilibrium point) cannot be achieved by using current control parameter A natural question is that how to systematically tune the system parameters such that the system can reach an asymptotic stable equilibrium point, which will be dealt in the following subsections.
% subsection motivating_example (end)

\subsection{Systematic Scheme for Parameter Tuning} % (fold)
\begin{figure}[htb]
    \centering
    \includegraphics[width = 3in]{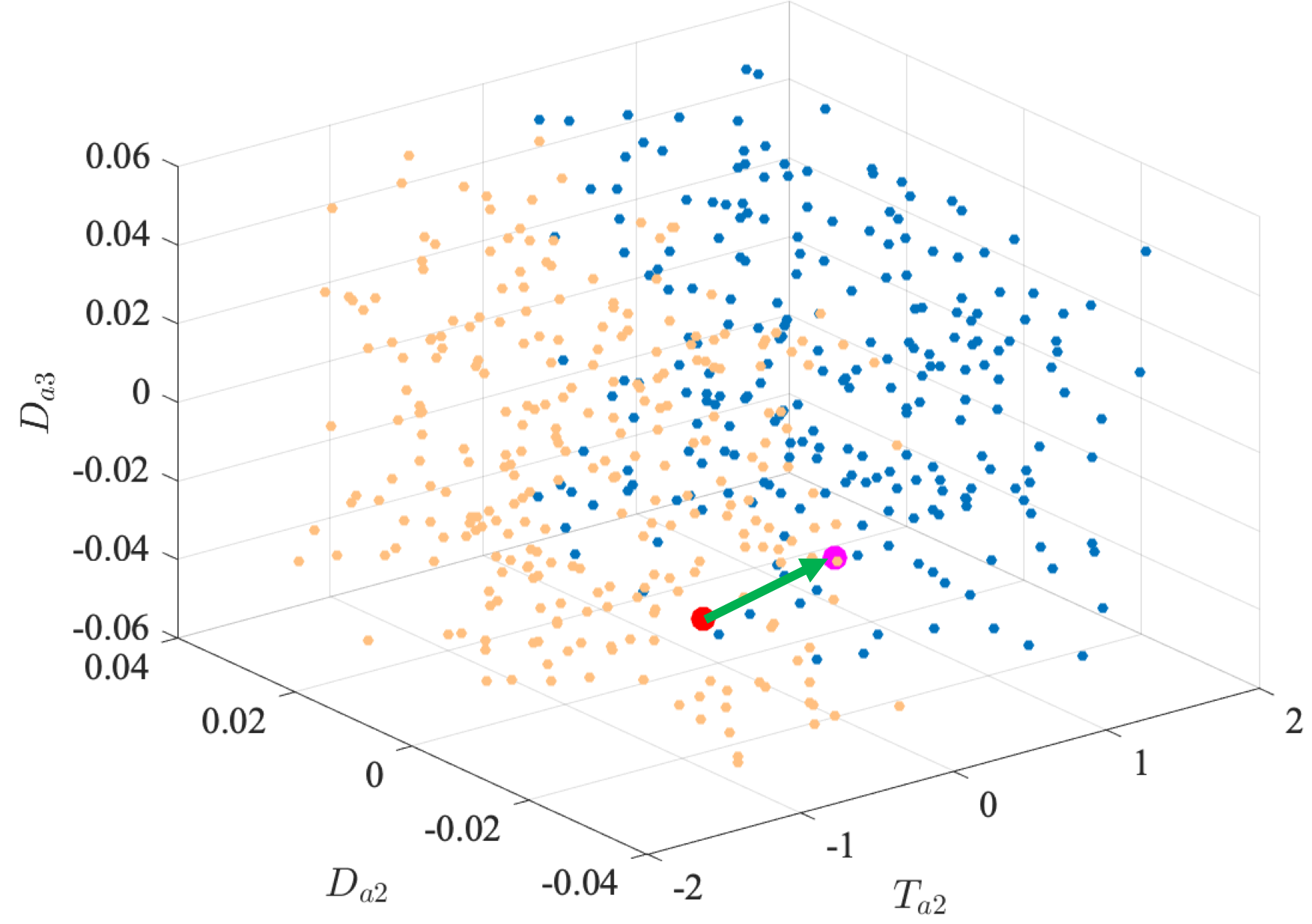}
    \caption{Visualization of desirable (blue), undesirable (orange), initial (red) and suggested (pink) parameters.}
    \label{fig:scatter_plot}
\end{figure}

\begin{figure}[htb]
    \centering
    \includegraphics[width = 2in]{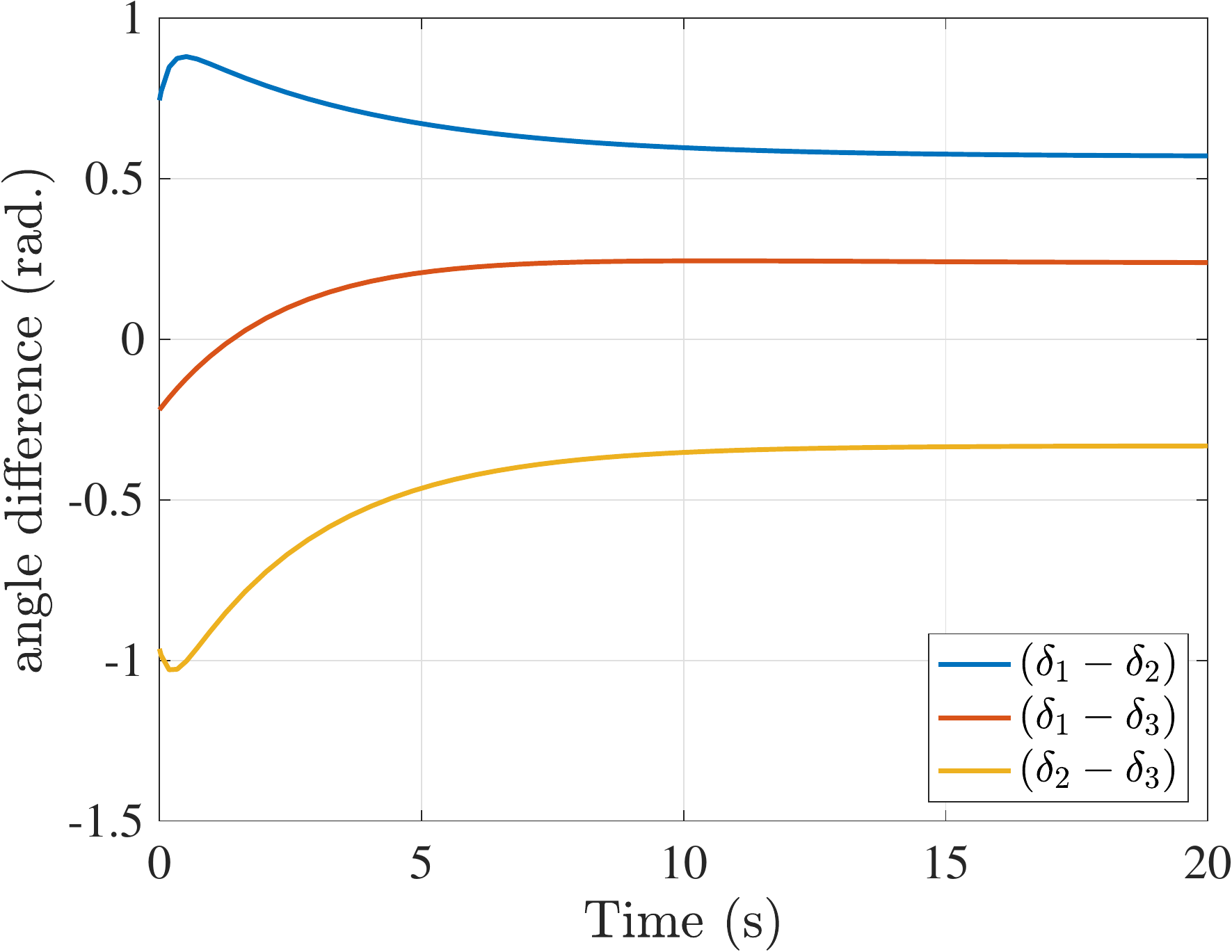}
    \caption{The evolution of voltage angle differences after the microgrid $4$ enters the islanding mode using the suggested parameters.
}
    \label{fig:angle_stable}
\end{figure}

\label{sub:systematic_scheme_for_parameter_tuning}
Assume that the adjustable parameters are $T_{\text{a}2}$, $D_{\text{a}2}$, and $D_{\text{a}3}$ in the test system. Algorithm \ref{alg:2} can be employed to systematically modify these tunable parameters. In Algorithm \ref{alg:2}, $N=500$,  $l_V = 4$, $l_{\mathbf{s}1}=2$, $l_{\mathbf{s}2}=2$, $l_{\sigma_1}=4$, and $l_{\sigma_2}=5$. The adjustable parameters are allowed to vary from $-200\%$ to $200\%$ of their original values, i.e., $\gamma_i'=-2, \gamma_i''=2$ in \eqref{eq:randomized}. Algorithm \ref{alg:2} first searches parameters $\mathcal{S}$ leading to an asymptotic stable equilibrium point, which are visualized in Figure \ref{fig:scatter_plot}. Then, the parameter combination closest to the initial parameter combination is considered as a suggestion for parameter update. The suggested parameters returned by Algorithm \ref{alg:2} are $T_{\text{a}2}=0.183$, $D_{\text{a}2}=-0.015$, and $D_{\text{a}3}=-0.025$. As shown in Fig \ref{fig:angle_stable}, with the parameters suggested by Algorithm \ref{alg:2}, the pre-designed equilibrium point is accomplished after MG $4$ enters the islanding mode due to disaster.
% subsection systematic_scheme_for_parameter_tuning (end)

\section{Conclusion}
\label{sec:conclusion}
In this paper, we propose a holistic framework for parameter coordination of a power electronic-interfaced microgrid interconnection against natural disasters. We identify a transient stability issue in a microgrid interconnection.
A novel transient stability assessment algorithm is designed based on SOS programming. Embedded with the stability assessment framework, a holistic framework is proposed for the purpose of systematically coordinating parameters such that post-disaster equilibrium points of microgrid interconnections are asymptotically stable. The efficacy of the proposed framework is tested in a four-microgrid interconnection. Future work will investigate the conservativeness of the SOS-based stability assessment algorithm and extend the framework to different control layers microgrid interconnections.

\bibliographystyle{IEEEtran}
\bibliography{confref.bib}

\end{document}